\documentclass[12pt]{article}   	
\usepackage{graphicx}				
\usepackage{amssymb}
\usepackage{amsmath}         
\usepackage{indentfirst}
\usepackage{enumerate}      
\usepackage{footmisc}          

\usepackage{graphicx}
\usepackage{subfigure} 

\usepackage{xcolor}

\usepackage[superscript]{cite}    

\begin{document} 

\title{Binary Blends of Diblock Copolymers: An Efficient Route to Complex Spherical Packing Phases}
\date{}

\author{Jiayu Xie, Yu Li, An-Chang Shi*\\
Department of Physics and Astronomy, McMaster University\\
1280 Main Street West, Hamilton, Ontario Canada L8S 4M1\\
Email Address: shi@mcmaster.ca}

\maketitle

\begin{abstract}
The phase behaviour of binary blends composed of A$_1$B$_1$ and A$_2$B$_2$ diblock copolymers is systematically studied using the polymeric self-consistent field theory, focusing on the formation and relative stability of various spherical packing phases. The results are summarized in a set of phase diagrams covering a large phase space of the system. Besides the commonly observed body-centered-cubic (BCC) phase, complex spherical packing phases including the Frank-Kasper A15 and $\sigma$ and the Laves C14 and C15 phases could be stabilized by the addition of longer A$_2$B$_2$-copolymers to asymmetric A$_1$B$_1$-copolymers. Stabilizing the complex spherical packing phases requires that the added A$_2$B$_2$-copolymers have a longer A-block and an overall chain length at least comparable to the host copolymer chains. A detailed analysis of the block distributions reveals the existence of inter- and intra-domain segregation of different copolymers, which depends sensitively on the copolymer length ratio and composition. The predicted phase behaviours of the A$_1$B$_1$/A$_2$B$_2$ diblock copolymer blends are in good agreement with available experimental and theoretical results. The study demonstrated that binary blends of diblock copolymers provide an efficient route to regulate the emergence and stability of complex spherical packing phases.
\end{abstract}

\section{Introduction}

Block copolymers are macromolecules composed of two or more chemically distinct sub-chains or blocks \cite{Bates:2017ir}. Due to the intrinsic frustration originated from a competition between the monomer-monomer interactions and chain connectivity, block copolymers tend to self-assemble into mesoscopic polymeric domains of various shapes, loosely classified as lamellae, cylinders and spheres \cite{shi2021frustration}. In block copolymer melts or concentrated block copolymer solutions, the packing of these domains leads to the formation of various ordered phases or mesocrystals. The formation and relative stability of these ordered phases have been an actively researched topic attracting sustained attention \cite{Bates:2017ir}. In particular, the emergence of complex spherical packing phases such as the Frank-Kasper phases has attracted tremendous attention in recent years \cite{lee2010discovery, schulze2017conformational, xie2014sigma,  liu2016stabilizing, li2017nonclassical, reddy2018stable, BatesMW2020, mueller2020emergence, lindsay2020a15,lindsay2021blends}.

The Frank-Kasper (FK) phases are a class of complex spherical packing phases initially discovered in metallic alloys. An important feature of the FK phases is the existence of at least two non-equivalent particles or Wigner-Seitz cells (WSCs) in the unit cell of the lattice, which compactly pack together in a complex manner to fill the space \cite{frank1958complex, frank1959complex}. Besides hard condensed matter systems \cite{deGraef2012}, the FK phases have been discovered in various soft matter systems, such as block copolymer melts and blends \cite{lee2010discovery, mueller2020emergence, lindsay2020a15,lindsay2021blends}, surfactant solutions \cite{kim2017low, jayaraman2018counterion, baez2018micellar} and giant molecules \cite{yue2016geometry, su2020constituent}. In the case of polymeric systems containing block copolymers, the FK $\sigma$ phase was first discovered in AB diblock copolymer melts \cite{lee2010discovery}. Since then a large number of studies have been carried out to understand the emergence of complex spherical packing phases in polymeric systems containing block copolymers both experimentally \cite{schulze2017conformational, lewis2018role, mueller2020emergence,lindsay2020a15,lindsay2021blends} and theoretically\cite{xie2014sigma, liu2016stabilizing, li2017nonclassical, reddy2018stable}. 

Different from hard condensed matter, in soft matter systems such as block copolymer melts, the particles or the spherical domains are deformable. Due to the broken spherical symmetry in a crystalline lattice, the packed soft spheres tend to deform towards the polyhedral shape of the WSCs by which they are enclosed in order to maintain a uniform monomer density. Such distortion would inevitably increase the free energy of the polymeric domains, which prefers their native spherical shapes. Thus, phases with higher average sphericity of the WSCs would be preferred when the domain is large enough to induce severe distortion. Based on this argument, the FK phases, which have higher average sphericity than the classical body-centered cubic (BCC) and close-packed (HCP or FCC) phases, could become stable if the spherical domain can be enlarged and, at the same time, the transition to cylindrical domains could be prevented \cite{lee2014sphericity,reddy2018stable}. 

A key feature of the complex spherical packing phases is the formation of large spherical domains. There are a number of routes to regulate the size of the polymeric domains \cite{li2017nonclassical,shi2021frustration}. Introducing conformational or configurational asymmetry into diblock copolymers is an effective approach to enlarge spherical domains. Indeed, a theoretical study based on the polymeric self-consistent field theory (SCFT) showed that the FK A15 and $\sigma$ phases could be stabilized in conformationally asymmetric linear AB-diblock copolymers and configurationally asymmetric $\text{AB}_{4}$ miktoarm copolymers \cite{xie2014sigma}. This theoretical prediction has been confirmed in experiments on conformationally asymmetric diblock copolymers \cite{schulze2017conformational} and on miktoarm AB block copolymers \cite{BatesMW2020}. Moreover, a recent SCFT study by Qiang {\em et al.} \cite{qiang2020stabilizing} demonstrated that by employing specially designed dendritic AB-type block copolymers, the spherical domains could be maintained up to $f_{A}\sim0.7$ resulting in large windows of FK A15 and $\sigma$ phases in the phase diagram. These studies offer a good understanding of the formation of FK phases in single component systems composed of asymmetric block copolymers.

Another feature of the complex spherical packing phases is that the sizes of the non-equivalent particles or polymeric domains are different. For example, there are two different types of particles in the A15 phase and five different particles in the $\sigma$ phase. These non-equivalent polymeric domains have different volumes, which is in contrast to the classical BCC, FCC and HCP phases having only one type of domains. This feature is more pronounced for the Laves C14 and C15 phases, which have spherical domains with much larger volume differences when compared with the A15 and $\sigma$ phases. The fact that the Laves phases have not been found to be stable in diblock copolymer melts could be attributed to the fact that their formation requires considerable volume exchange among distinct domains, which is not favoured in single-component systems. From this perspective, one mechanism to stabilize the complex spherical packing phases is to regulate the sizes of the polymeric domains, which could be effectively accomplished by mixing another species into the system \cite{li2017nonclassical,shi2021frustration}.

Indeed, experimental and theoretical studies have suggested that blending different components together could provide an efficient route to regulate the domain sizes thus stabilizing the complex spherical packing phases. The simplest blending system is obtained by mixing A-homopolymers with sphere-forming AB diblock copolymers. The added A-homopolymers would be localized in the central region, or the core, of the spherical domains resulting in larger spheres. At the same time, differential segregation of the A-homopolymers would lead to the formation of spherical domains with different sizes. The combined effects could stabilize the formation of complex spherical packing phases. This route has been demonstrated experimentally by Mueller {\em et al.} showing that the FK $\sigma$, Laves C14 and C15 phases could become stable phases in AB/A binary blends in the dry brush regime \cite{mueller2020emergence}. Moreover, the local segregation of the A-homopolymers has been illustrated by SCFT calculations for AB/A \cite{cheong2020symmetry,xie2021giant} and $\text{AB}_4$/A \cite{zhao2019laves} binary blends. Another blending platform is mixing AB diblock copolymers with different lengths and compositions. The formation of complex spherical packing phases in block copolymer blends has been examined theoretically \cite{liu2016stabilizing,kim2018origins}, predicting that complex spherical packing phases, including the A15, $\sigma$, C14 and C15 phases, could be formed by mixing $\text{A}_1\text{B}_1$ and $\text{A}_2\text{B}_2$ diblock copolymers. In agreement with the theoretical predictions, these complex spherical packing phases have been observed in recent experiments \cite{lindsay2020a15,lindsay2021blends} on binary blends of $\text{A}_1\text{B}_1/\text{A}_2\text{B}_2$ diblock copolymers. These previous experimental and theoretical studies have shown that the binary blends composed of $\text{A}_1\text{B}_1/\text{A}_2\text{B}_2$ diblock copolymers provide a flexible platform to stabilize complex spherical packing phases, as well as a useful model system to study the mechanism of the formation of these complex phases. The fundamental difference between the addition of A-homopolymers and AB diblock copolymers to an AB diblock copolymer melt is that the added AB diblock copolymers could be localized inside one domain or at the AB-interfaces. As such, the added AB diblock copolymers could act as filler component similar to A-homopolymers {\it and} as co-surfactants which could modify the property of the AB-interfaces. A synergetic interplay of these two functions could lead to a much enhanced effect on the stabilization of the complex spherical packing phases. 

The self-assembly of binary blends of $\text{A}_1\text{B}_1/\text{A}_2\text{B}_2$ diblock copolymers has been investigated experimentally \cite{hashimoto1994blends,Yamaguchi:1997p20,Yamaguchi:2000p1200,Yamaguchi:2001p185,hashimoto2002blends,hashimoto2007blends,lindsay2020a15,lindsay2021blends} and theoretically \cite{SHI:1994p2,Shi:1995tc,Matsen:1995p1143,Wu:2010p1978,Wu:2011p2198,Lai2021blends,liu2016stabilizing,kim2018origins} in the past years. It has been well established that mixing $\text{A}_1\text{B}_1$ and $\text{A}_2\text{B}_2$ diblock copolymers could lead to a rich phase behaviour including macroscopic phase separation and the emergence of new phases. In particular, previous experiments \cite{lindsay2020a15,lindsay2021blends} and theory \cite{liu2016stabilizing,kim2018origins} have laid a foundation of the the self-assembly of complex spherical packing phases in $\text{A}_1\text{B}_1/\text{A}_2\text{B}_2$ diblock copolymer blends. However, a complete set of phase diagrams covering a large phase space of the system and a detailed investigation of the effect of different molecular parameters are still lacking. Therefore, it is desirable to carry out a systematic study on the binary $\text{A}_1\text{B}_1/\text{A}_2\text{B}_2$ diblock copolymer blends covering a large phase space. In the current work, we fill this gap by carrying out a comprehensive study of $\text{A}_1\text{B}_1/\text{A}_2\text{B}_2$ blends using the polymeric self-consistent field theory. We will mainly focus on the effects of three parameters: (1) the concentration of the $\text{A}_2\text{B}_2$-copolymers $\phi_2$; (2) the composition of the $\text{A}_2\text{B}_2$-copolymers $f_2$; and (3) the length ratio between the $\text{A}_1\text{B}_1$- and $\text{A}_2\text{B}_2$-copolymers $\gamma$. A set of phase diagrams in the $\phi_{2}-\chi{N}$ plane with different values of $f_2$ and $\gamma$ are constructed and presented. The phase diagrams cover a large range of $\phi_2$ from 0 to 1 and $\chi{N}$ from 0 to 40. These phase diagrams span a large region of the phase space and give a systematic overview of the phase behaviour of the binary blends. The predicted phase transition sequences could be used to make direct comparison with experimental phase diagrams in the concentration v.s. temperature plane. Furthermore, we perform a detailed analysis of the effects of different molecular parameters on the inter- and intra-domain segregation of the copolymers. Our results provide a comprehensive picture of the phase behaviour of binary $\text{A}_1\text{B}_1/\text{A}_2\text{B}_2$ diblock copolymer blends, thus shedding light on the mechanisms of the stabilization of complex spherical packing phases.

\section{Theoretical model}
We consider an incompressible binary blend composed of linear $\text{A}_1\text{B}_1$ and $\text{A}_2\text{B}_2$ diblock copolymers in a volume $V$. Specifically, the model system contains $n_1$ $\text{A}_1\text{B}_1$ diblock copolymer chains and $n_2$ $\text{A}_2\text{B}_2$-- diblock copolymer chains. The degree of polymerization of the $\text{A}_1\text{B}_1$- and $\text{A}_2\text{B}_2$-copolymers is $N_1={\gamma_1}N=N$ and $N_2={\gamma_2}N=\gamma{N}$ ($\gamma_1=1, \gamma_2=\gamma$), respectively. The volume fraction of the A-blocks for the two copolymers is $f_1=N_{1A}/N_1$ and $f_2=N_{2A}/N_2$, respectively. We assume a uniform segment density $\rho_{\text{A},0}=\rho_{\text{B},0}=\rho_0$ such that ${\rho_0}V=n_{1}N+n_{2}{\gamma}N$ according to the incompressibility condition. The average concentrations of the $\text{A}_1\text{B}_1$- and $\text{A}_2\text{B}_2$-copolymers are given by, $\phi_1=\frac{n_{1}{\gamma_1}N}{{\rho_0}V}$ and $\phi_2=\frac{n_{2}{\gamma_2}N}{{\rho_0}V}=1-\phi_{1}$, respectively.
Furthermore, we denote the Kuhn length of the A- and B-segments by $b_\text{A}$ and $b_\text{B}$, which can take different values to model chains with different stiffness or conformational asymmetry.

In order to consider two phase coexistence, it is convenient to formulate the theory in the grand canonical ensemble in which the thermodynamic parameters are the chemical potentials of the two copolymers, $\mu_1$ and $\mu_2$. Because the system is assumed to be incompressible, these two chemical potentials are not independent and can be chosen as $\mu_{1}=0$ and $\mu_{2}=\mu$ for convenience. Within the mean-field approximation, the grand potential density of the system can be expressed as \cite{fredrickson2006equilibrium,shi2017variational},
\begin{align}
\begin{split}
\frac{N\Phi}{\rho_{0}{V}k_{\text{B}}T}=&-Q_{1}-\text{e}^{\frac{\mu}{k_\text{B}T}}Q_{2}+\frac{1}{V}\int{d\vec{r}}\left\{\chi{N}\phi_\text{A}(\vec{r})\phi_\text{B}(\vec{r})\right.\\
&\left.-\omega_\text{A}(\vec{r})\phi_\text{A}(\vec{r})-\omega_\text{B}(\vec{r})\phi_\text{B}(\vec{r})-\eta(\vec{r})\left[1-\phi_\text{A}(\vec{r})-\phi_\text{B}(\vec{r})\right]\right\}.
\end{split}
\end{align}
By minimizing the grand potential with respect to the density and fields, we obtain the SCFT equations,
\begin{align}\label{SCFT}
\left\{
\begin{aligned}
&\omega_{\text{A}}(\vec{r})=\chi{N}\phi_{\text{B}}(\vec{r})+\eta(\vec{r}),\\
&\omega_{\text{B}}(\vec{r})=\chi{N}\phi_{\text{A}}(\vec{r})+\eta(\vec{r}),\\
&\phi_{\text{A}}(\vec{r})=\sum_{i=1}^{2}\text{e}^{\frac{\mu_i}{k_\text{B}T}}\int_{0}^{{\gamma_i}f_{i}}ds\,q_{i}(s,\vec{r})q_{i}^{\dag}(s,\vec{r}),\\
&\phi_{\text{B}}(\vec{r})=\sum_{i=1}^{2}\text{e}^{\frac{\mu_i}{k_\text{B}T}}\int_{\gamma_i f_i}^{{\gamma_i}}ds\,q_{i}(s,\vec{r})q_{i}^{\dag}(s,\vec{r}),\\
&\phi_{\text{A}}(\vec{r})+\phi_{\text{B}}(\vec{r})=1,
\end{aligned}
\right.
\end{align}
where we have used $N$ as the scale of the polymer arc-length.

In Eqs.~\ref{SCFT}, the forward propagators $q_{i}(s,\vec{r})$ and backward propagators $q_{i}^{\dag}(s,\vec{r})$ ($i=1,2$) are obtained by solving the modified diffusion equations,
\begin{align} \label{MDEs}
\begin{split}
\frac{\partial}{\partial{s}}q_{i}(s,\vec{r})=&\epsilon_i^2(s)\nabla^{2}q_{i}(s,\vec{r})-\omega_{i}(s,\vec{r})q_{i}(s,\vec{r}),\\
-\frac{\partial}{\partial{s}}q_{i}^{\dag}(s,\vec{r})=&\epsilon_{i}^2(s)\nabla^{2}q_{i}^{\dag}(s,\vec{r})-\omega_{i}(s,\vec{r})q_{i}^{\dag}(s,\vec{r}),
\end{split}
\end{align}
where $s\in[0,\gamma_i]$ for the $\text{A}_i\text{B}_i$-copolymers ($i=1,2$). The function $\epsilon_{i}^2(s)$ represents the conformational asymmetry parameter, which is defined by $\epsilon_{i}^2(s)=(\rho_{\text{A},0}b_\text{A}^2)/(\rho_{\text{B},0}b_\text{B}^2)=b_\text{A}^2/b_\text{B}^2$ for $s\in[0,{\gamma_i}f_i]$ and $\epsilon_{i}^2(s)=1$ for $s\in[{\gamma_i}f_i,\gamma_i]$. The self-consistent fields $\omega_{i}(s,\vec{r})$ for the $\text{A}_i\text{B}_i$ copolymers are given by, $\omega_{i}(s,\vec{r})=\omega_{\text{A}}(\vec{r})$ when $s\in[0,\gamma_{i}f_i]$ and $\omega_{i}(s,\vec{r})=\omega_{\text{B}}(\vec{r})$ when $s\in[\gamma_{i}f_i,\gamma_{i}]$. 
The initial conditions of the propagators are specified by $q_{i}(0,\vec{r})=1$ and $q_{i}^{\dag}(\gamma_{i},\vec{r})=1$.

The single chain partition functions are obtained from the solutions of the propagators by,
\begin{align}
Q_{i}=\frac{1}{V}{\int}d\vec{r}q_{i}(\vec{r},\gamma_i).
\end{align}
Once the SCFT equations (Eqs.~\ref{SCFT}) are solved, the average concentrations of the two components can be simply calculated by,
\begin{gather}
\phi_1=Q_1,\, \phi_2=1-\phi_1.
\end{gather}

In many cases it is advantageous to specify the concentration of each components, or equivalently the numbers of polymer chains, in the system explicitly. In this case it is convenient to work in the canonical ensemble, where the concentration of the $\text{A}_i\text{B}_i$-copolymers ($\phi_i$, $i=1,2$) are the thermodynamic control parameters. Within the mean-field theory, the Helmholtz free energy density of the system is expressed as \cite{fredrickson2006equilibrium,shi2017variational},
\begin{align}
\begin{split}
\frac{NF}{\rho_{0}{V}k_{\text{B}}T}=&-\phi_{1}{\ln}\frac{Q_{1}}{\phi_1}-\frac{\phi_{2}}{\gamma}{\ln}\frac{Q_{2}}{\phi_2}+\frac{1}{V}\int{d\vec{r}}\left\{\chi{N}\phi_\text{A}(\vec{r})\phi_\text{B}(\vec{r})\right.\\
&\left.-\omega_\text{A}(\vec{r})\phi_\text{A}(\vec{r})-\omega_\text{B}(\vec{r})\phi_\text{B}(\vec{r})-\eta(\vec{r})\left[1-\phi_\text{A}(\vec{r})-\phi_\text{B}(\vec{r})\right]\right\}.
\end{split}
\end{align}
Minimization of this free energy with respect to the densities and fields results in the following SCFT equations,
\begin{align}
\left\{
\begin{aligned}
&\omega_{\text{A}}(\vec{r})=\chi{N}\phi_{\text{B}}(\vec{r})+\eta(\vec{r}),\\
&\omega_{\text{B}}(\vec{r})=\chi{N}\phi_{\text{A}}(\vec{r})+\eta(\vec{r}),\\
&\phi_{\text{A}}(\vec{r})=\sum_{i=1}^{2} \frac{\phi_i}{\gamma_i Q_i} \int_{0}^{{\gamma_i}f_{i}}ds\,q_{i}(s,\vec{r})q_{i}^{\dag}(s,\vec{r}),\\
&\phi_{\text{B}}(\vec{r})=\sum_{i=1}^{2} \frac{\phi_i}{\gamma_i Q_i} \int_{\gamma_i f_i}^{{\gamma_i}}ds\,q_{i}(s,\vec{r})q_{i}^{\dag}(s,\vec{r}),\\
&\phi_{\text{A}}(\vec{r})+\phi_{\text{B}}(\vec{r})=1,
\end{aligned}
\right.
\end{align}
where the propagators are computed by the same modified diffusion equations in Eqs.~\ref{MDEs}.
\par

The SCFT equations are a set of coupled nonlinear and non-local equations. For most of the cases, solutions of the SCFT equations should be obtained numerically. We employ the pseudo-spectral method to solve the modified diffusion equations \cite{rasmussen2002improved,tzeremes2002efficient}. Moreover, the variable-cell Anderson mixing scheme \cite{thompson2004improved,arora2017accelerating} is used to speed up the self-consistent iteration and minimize the free energy or grand-potential with respect to the unit-cell dimensions simultaneously. Starting from specific initial configurations, a large number of solutions corresponding to different ordered phases can be obtained. The relative stability of these different phases is determined by comparing their free energy or grand potential density. The phase boundaries between two phases are found by locating the intersections between their thermodynamic potentials.

\section{Results and Discussion}
\subsection{Phase Behaviour}
For the model binary $\text{A}_1\text{B}_1/\text{A}_2\text{B}_2$ diblock copolymer blends, their phase behaviour is controlled by six independent parameters, namely, the A-block volume fractions $f_1$ and $f_2$, molecular weight ratio $\gamma$, conformational asymmetry parameter $\epsilon$, copolymer concentration $\phi_2$, and segregation strength $\chi{N}$. In the current study we consider conformationally symmetric diblock copolymers with a fixed $\epsilon=1$ because we are focusing on the effect of blending rather than conformational asymmetry. It is noted that the effect of conformational asymmetry has been studied extensively for AB-diblock copolymer melts by theory \cite{xie2014sigma} and experiments \cite{schulze2017conformational}. In order to obtain a comprehensive overview of the phase behaviour of the $\text{A}_1\text{B}_1/\text{A}_2\text{B}_2$ diblock copolymer blends, we will construct two sets of phase diagrams in the $\phi_{2}-\chi{N}$ plane, highlighting the effect of $f_2$ and $\gamma$, respectively. There are two main benefits of this construction. Firstly, different from previous works \cite{liu2016stabilizing, kim2018origins}, where the length of either the majority or the minority block of the longer chain are fixed relative to the shorter chain so that $f_2$ and $\gamma$ are coupled, we choose to decouple the effects of these two parameters to provide an understanding of the phase behaviours from different perspectives. Specifically, we are able to examine the effects of $f_2$ and $\gamma$ separately to reveal the synergetic effects of $f_2$ and $\gamma$ on the stabilization of the complex spherical packing phases. Secondly, we choose to construct the phase diagrams in the $\phi_{2}-\chi{N}$ plane so that the results can be compared directly with experimental observations \cite{lindsay2020a15,lindsay2021blends}.

\begin{figure*}[htbp]
\centering
\includegraphics[width=7cm]{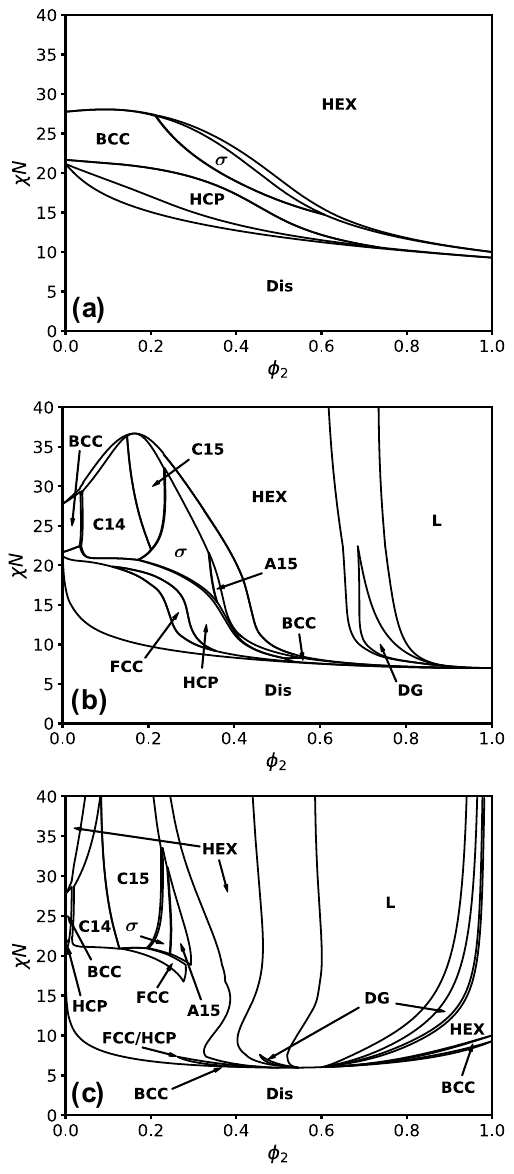}
\caption{Phase diagram in the $\phi_2-\chi{N}$ plane for (a) $f_{2}=0.3$, (b) $f_{2}=0.5$ and (c) $f_{2}=0.7$ with fixed $f_{1}=0.2$ and $\gamma=1.5$. The unlabeled regions are two-phase coexistence regions where two adjacent single phases are connected by a horizontal tie-line. In (c), the FCC/HCP indicates that these two phases are degenerate within the accuracy of the SCFT calculations and the phases on the right hand side of L are the inverse phases.}
\label{PD1}
\end{figure*}

\begin{figure*}[htbp]
\centering
\includegraphics[width=7cm]{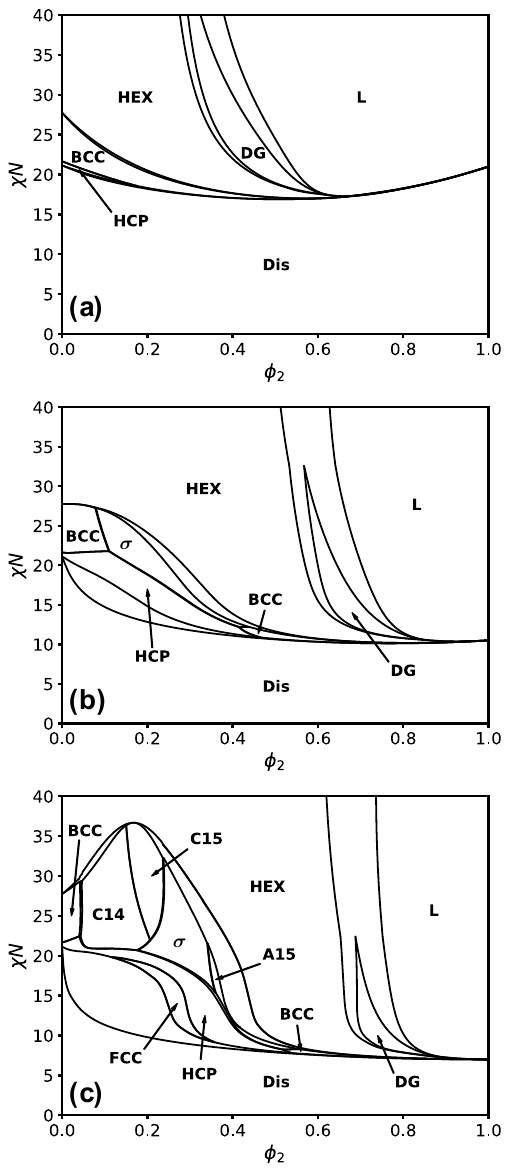}
\caption{Phase diagrams in the $\phi_2-\chi{N}$ plane for (a) $\gamma=0.5$, (b) $\gamma=1.0$ and (c) $\gamma=1.5$ with fixed $f_{1}=0.2$ and $f_{2}=0.5$. The unlabeled regions are the two-phase coexistence regions between two adjacent single phases. Note that Fig.\ref{PD2}(c) is identical to Fig.\ref{PD1}(b), which is reproduced here for easy comparison.}
\label{PD2}
\end{figure*}

Phase diagrams of the blends are constructed by comparing the free energy of various candidate phases including the disordered (Dis), lamellar (L), double gyroid (DG), hexagonally-packed cylinders (HEX), body-centered-cubic (BCC), face-centered-cubic (FCC), hexagonally close-packed spheres (HCP) as well as the FK A15, $\sigma$, Laves C14 and C15 phases. 
This choice of candidate phases is motivated by our knowledge of potential equilibrium phases obtained from previous experimental and theoretical studies. We have obtained SCFT solutions of several Frank-Kasper phases, such as the Z and H phases, beyond this list. However, these phases have not found to become stable phase within the parameter space explored in the current study.
The phase behaviours of the binary $\text{A}_1\text{B}_1$/$\text{A}_2\text{B}_2$ diblock copolymers are summarized in two sets of phase diagrams shown in Figures~\ref{PD1} and \ref{PD2}, respectively. The first set of phase diagrams (Figure~\ref{PD1}) are for the cases with $f_1=0.2$ and $\gamma=1.5$ and three typical values of $f_2=0.3$, $0.5$ and $0.7$. This choice of $f_2$ covers both symmetric case and two oppositely asymmetric cases for the $\text{A}_2\text{B}_2$-copolymers. The second set of phase diagrams (Figure~\ref{PD2}) are for the cases with $f_1=0.2$ and $f_2=0.5$ and three typical values of $\gamma=0.5$, $1.0$ and $1.5$. This choice of $\gamma$ covers the cases where the overall chain length of the $\text{A}_2\text{B}_2$-copolymers is shorter, equal and longer than that of the $\text{A}_1\text{B}_1$-copolymers. All phase diagrams are computed in the full range of $\phi_2$ from 0 to 1 and a large  range of $\chi{N}$ from 0 to 40. Taking together, these two sets of phase diagrams represent the phase behaviour of the model system in a large region of the phase space.

In the phase diagrams shown in Figures~\ref{PD1} and \ref{PD2}, the vertical lines at $\phi_2$=0 and 1 represent the phase behaviours of the neat $\text{A}_1\text{B}_1$ and $\text{A}_2\text{B}_2$ diblock copolymers, respectively. Because these two diblock copolymers are conformationally symmetric ($b_A=b_B$), the stable phases of the neat diblock copolymers are the classical phases. For example, the phase transition of the neat diblock copolymers follows the generic sequence of Dis $\rightarrow$ HCP $\rightarrow$ BCC $\rightarrow$ HEX $\rightarrow$ L for $f_i=0.2$ when the segregation strength $\chi{N}$ is increased. Upon the blending of the two diblock copolymers, new ordered stable phases including the complex spherical packing phases could emerge as stable phases. As the $\text{A}_2\text{B}_2$-copolymer concentration ($\phi_2$) is increased, the system undergoes a series of phase transitions from the neat, $f_1$-dependent, $\text{A}_1\text{B}_1$ phase to the neat, $f_2$-dependent, $\text{A}_2\text{B}_2$ phase. The phase transition sequence of the blends as $\phi_2$ changes from $0$ (neat $\text{A}_1\text{B}_1$) to $1$ (neat $\text{A}_2\text{B}_2$) depends sensitively on the segregation strength $\chi{N}$, the volume fractions of the A-blocks $f_i$, and the chain length ratio $\gamma$. In particular, the phase transition sequence could vary from no phase transition at all to a complex one involving up to a dozen ordered phases. It is noted that the SCFT calculations are within the framework of mean-field theory and the order-disorder transition boundary is expected to be modified by fluctuation effects. Nevertheless, we expect that the phase transition sequence at larger $\chi{N}$ are less affected by fluctuations.

The phase diagrams for $f_1=0.2$ and $\gamma=1.5$ shown in Figure~\ref{PD1} highlight the effects of varying $f_2$ on the phase behaviour of the $\text{A}_1\text{B}_1$/$\text{A}_2\text{B}_2$ diblock copolymer blends. For $f_1=0.2$, the neat $\text{A}_1\text{B}_1$ diblock copolymers undergoes phase transitions following the sequence of Dis $\rightarrow$ HCP $\rightarrow$ BCC $\rightarrow$ HEX as $\chi{N}$ is increased from $0$ to $40$. For $f_2=0.3$ that is similar to $f_1=0.2$, the neat $\text{A}_2\text{B}_2$ diblock copolymers exhibit a phase transition sequence of Dis $\rightarrow$ BCC $\rightarrow$ HEX as $\chi{N}$ is increased. In this case the phase behaviour of the $\text{A}_1\text{B}_1$/$\text{A}_2\text{B}_2$ blends (Figure~\ref{PD1}(a)) is relatively simple. At small ($\chi{N} \leqslant 10$) and large ($\chi{N} \geqslant 27$) values of $\chi{N}$, the addition of the $\text{A}_2\text{B}_2$ diblock copolymers does not induce any phase transitions. That is, the binary $\text{A}_1\text{B}_1$/$\text{A}_2\text{B}_2$ blends would be in the disordered phase at small $\chi{N}$ and the hexagonally-packed cylindrical phase at large $\chi{N}$, respectively. In the intermediate region ($10 \leqslant \chi{N} \leqslant 27$), order-to-order phase transitions could be induced with the addition of the $\text{A}_2\text{B}_2$ diblock copolymers. In particular, a small window of the FK $\sigma$ phase appears in the middle ($0.2 \leqslant \phi_2 \leqslant 0.6$) of the phase diagrams. The predicted phase transition sequence as a function of $\phi_2$ follows the generic sequence of Dis $\rightarrow$ HCP $\rightarrow$ BCC $\rightarrow$ $\sigma$ $\rightarrow$ HEX in which some of the phases could be missing for a given $\chi{N}$. Notably, although the two diblock copolymers have similar A-block volume fractions and there is no conformational asymmetry, the addition of the $\text{A}_2\text{B}_2$-copolymers could stabilize a complex spherical packing phase, {\it albeit} in a small region on the phase diagram, suggesting that even a modest difference in the A-block lengths is sufficient to stabilize the $\sigma$ phase.

The phase behaviour becomes much more richer for the cases of larger $f_2$, as shown in the phase diagrams for $f_2=0.5$ (Figure~\ref{PD1}(b)) and $f_2=0.7$ (Figure~\ref{PD1}(c)). When $f_2=0.5$, the neat ($\phi_2=1.0$) $\text{A}_2\text{B}_2$ diblock copolymers transitions from the disordered (Dis) phase to the lamellar (L) phase at $\chi{N_2}=\gamma\chi{N}=10.5$. The addition of this symmetric, lamella-forming diblock copolymers to the asymmetric $\text{A}_1\text{B}_1$ diblock copolymers results in an extremely rich phase behaviour as shown in Figure~\ref{PD1}(b). Comparing with the phase diagram for $f_2=0.3$ (Figure~\ref{PD1}(a)) containing 4 ordered phases (HCP, BCC, $\sigma$, and HEX), 10 ordered phases (HCP, FCC, BCC, C14, C15, $\sigma$, A15, HEX, DG and L) can become equilibrium phases in the phase diagram for $f_2=0.5$. It is amazing that a seemingly small change from $f_2=0.3$ to $f_2=0.5$ leads to such a drastic change of the phase behaviour. When $f_2$ is increased further to $f_2=0.7$ (Figure~\ref{PD1}(c)), the neat $\text{A}_2\text{B}_2$-copolymers form inverted BCC and HEX phases since the B-blocks are the domain-forming minority component. The phase diagram for the binary blends of diblock copolymers with $f_1=0.2$ and $f_2=0.7$ becomes even more complex as shown in Figure~\ref{PD1}(c). The formation of complex spherical packing phases persists in this case, except that the window of the complex spherical phases is pushed to smaller values of $\phi_2$. Furthermore, the region of the FK phases extends to higher $\chi{N}$ and the region of Laves phases grows at the cost of the $\sigma$ phase. Based on the phase diagrams shown in Figure~\ref{PD1}, we can conclude that the addition of the $\text{A}_2\text{B}_2$-copolymers with $f_2 \geqslant 0.5$ stabilizes the complex spherical packing phases, such that the Frank-Kasper $\sigma$ and A15 phases and the Laves C14 and C15 phases appear as equilibrium phases in the phase diagrams. This prediction provides an efficient route to obtain these complex spherical packing phases via blending of two diblock copolymers.

Besides the volume fraction of the A-blocks $f_2$, the chain length or the degree of polymerization ($N_2=\gamma{N}$) of the $\text{A}_2\text{B}_2$ diblock copolymers could also have a large effect on the phase behaviour of the binary blends. The effect of varying $\gamma$ on the phase behaviour of binary $\text{A}_1\text{B}_1$/$\text{A}_2\text{B}_2$ diblock copolymer blends is revealed in the phase diagrams shown in Figure~\ref{PD2} for the cases with $f_1=0.2$ and $f_2=0.5$. This set of phase diagrams clearly demonstrate the importance of the chain length of the added $\text{A}_2\text{B}_2$-copolymers. When the $\text{A}_2\text{B}_2$-copolymers are shorter than the $\text{A}_1\text{B}_1$-copolymers as exemplified by $\gamma=0.5$, the phase behaviour is relatively simple (Figure~\ref{PD2}(a)). In particular, the addition of short $\text{A}_2\text{B}_2$-copolymers does not stabilize the complex spherical packing phases and the order-to-order phase transition follows the generic sequence of HCP $\rightarrow$ BCC $\rightarrow$ HEX $\rightarrow$ DG $\rightarrow$ L (Figure~\ref{PD2}(a)). When the chain length of the $\text{A}_2\text{B}_2$-copolymers is increased to be the same as that of the $\text{A}_1\text{B}_1$-copolymers ($\gamma=1.0$), a small window of the FK $\sigma$ phase appears along the BCC/HEX phase boundary (Figure~\ref{PD2}(b)). Further increasing $\gamma$ from $\gamma=1$ to $\gamma=1.5$ results in the rich phase behaviour containing a set of 10 ordered phases including the complex spherical packing phases as shown in Figure~\ref{PD2}(c) that is the same as Figure~\ref{PD1}(b). An important conclusion from these theoretical results is that, in order to stabilize the complex spherical packing phases in the $\text{A}_1\text{B}_1$/$\text{A}_2\text{B}_2$ binary blends, sufficiently large values of $f_2$ {\em and} $\gamma$ are required.  As shall be presented in the next section, the reason of this behaviour is that the additive $\text{A}_2\text{B}_2$-copolymers with much shorter chain length than the host $\text{A}_1\text{B}_1$-copolymers tend to segregate at the AB-interface, which does not affect the packing of the A-blocks at the centre of the domains.

It is interesting to compare the phase behaviours presented in Figures~\ref{PD1} and \ref{PD2} to previous theoretical and experimental results. A set of theoretical phase diagrams of binary $\text{A}_1\text{B}_1$/$\text{A}_2\text{B}_2$ diblock copolymer blends have been constructed by Wu {\it et al.} \cite{Wu:2010p1978,Wu:2011p2198}. Because the complex spherical packing phases were not included in these previous studies, a detailed comparison of the phase boundaries is not straightforward. Nevertheless, the overall phase behaviour obtained in these previous studies is consistent with the current results summarized in Figures~\ref{PD1} and \ref{PD2}. In particular, both sets of phase diagrams exhibit the trend of the expansion of the HEX phase and the shrinking of the L phase when $\gamma$ is increased. In a more recent SCFT study \cite{liu2016stabilizing,kim2018origins}, it was predicted that the FK A15 and $\sigma$ phases and the Laves C14 and C15 phases could be stabilized in binary $\text{A}_1\text{B}_1$/$\text{A}_2\text{B}_2$ diblock copolymer blends. Phase diagrams in the $\phi_2-\gamma$ plane were obtained for a fixed $\chi{N}=40$ and two sets of $f_1=0.15$ and 0.45, whereas the value of $f_2$ varies with $\gamma$ as $f_2=0.85/\gamma$. Although the phase diagrams in the current work are provided in a different phase plane, the phase behaviours obtained from these two studies are completely consistent. It should be emphasized that the phase diagrams shown in Figures~\ref{PD1} and \ref{PD2} cover an unprecedented large phase space of the system, thus provide more information about the phase behaviour of the $\text{A}_1\text{B}_1$/$\text{A}_2\text{B}_2$ blends. 

Experimentally, the formation of complex spherical packing phases in binary $\text{A}_1\text{B}_1$/$\text{A}_2\text{B}_2$ diblock copolymer blends has been examined by Lindsay {\em et al.}\cite{lindsay2020a15,lindsay2021blends}. These authors observed a number of complex spherical packing phases including A15, $\sigma$, C14, C15 and a quasicrystal. The overall observed phase behaviour from these experiments is consistent with the theoretical predictions. In particular, the experiments reported by Lindsay {\em et al.} \cite{lindsay2020a15} revealed a phase transition sequence from $\sigma$ $\rightarrow$ A15 $\rightarrow$ HEX as $\phi_2$ is increased from 0.25 to 0.5. The same phase transition sequence is found in the phase diagram shown in Figure~\ref{PD1}(b), which has similar molecular parameters as that of the experiments, along the path at $\chi{N}\approx{20}$ with $\phi_2$ varies from 0.3 to 0.5. This good agreement between theory and experiment is very encouraging. 

\subsection{Segregation of Diblock Copolymers}

The phase diagrams presented above clearly reveal that the addition of a second $\text{A}_2\text{B}_2$ diblock copolymer with a larger A-volume fraction and longer chain length to sphere-forming $\text{A}_1\text{B}_1$ diblock copolymers provides an efficient route to stabilize complex spherical packing phases. It is desirable to understand the mechanisms favouring the formation of these complex ordered phases in the blends. The complex spherical packing phases are characterized by polymeric domains with different sizes and shapes. The formation of these complex structures would be favoured if the free energy cost of deforming the polymeric domains is reduced. As proposed by Liu {\em et al.} \cite{liu2016stabilizing}, there are two possible mechanisms stabilizing the complex spherical packing phases in binary $\text{A}_1\text{B}_1$/$\text{A}_2\text{B}_2$ diblock copolymer  blends, involving the segregation of the $\text{A}_1\text{B}_1$- and $\text{A}_2\text{B}_2$-copolymers among different polymeric domains (inter-domain segregation) and within each domain (intra-domain segregation).The inter- and intra-domain segregation of the copolymers provides mechanisms to regulate the size {\em and} shape of the spherical domains. In this subsection, we provide a detailed analysis of the two mechanisms based on a number of quantities extracted from the SCFT results. Additionally, we will demonstrate how the local copolymer segregation is affected by the molecular parameters, {\em i.e.} $\phi_2$, $f_2$ and $\gamma$.

\subsubsection{Inter-domain Segregation}

\begin{figure*}[htbp]
\centering
\includegraphics[width=12cm]{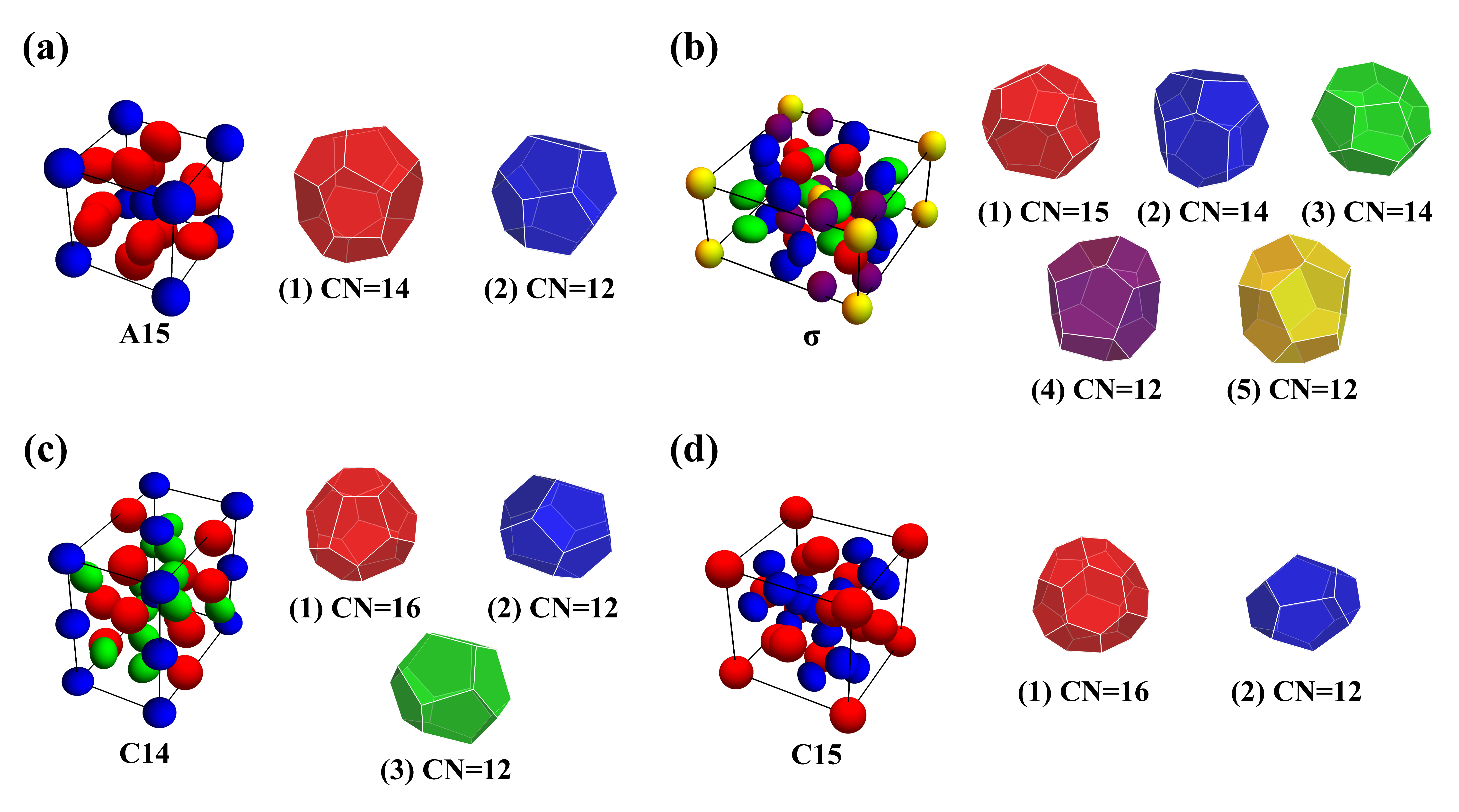}
\caption{The unit cell of (a) A15, (b) $\sigma$, (c) C14 and (d) C15 phases. Each type of WSC is plotted along with its coordination number.}
\label{FK}
\end{figure*}

The structure of the spherical phases could be regarded as the close packing of the space by unit cells. One particularly useful method to describe the packing pattern is to partition the space by closely-packed Wigner-Seitz cells (WSCs), such that each WSC encloses one minority A-domain or one ``particle". For the classical spherical packing phases (HCP, FCC and BCC), all the particles are symmetrically equivalent, so that there is only one type of WSC with the same volume or equivalently the same number of copolymers contained in each WSC. In contrast, the complex spherical packing phases such as the Frank-Kasper and Laves phases have at least two types of non-equivalent WSCs as illustrated in Figure~\ref{FK} \cite{deGraef2012} . These non-equivalent WSCs would naturally have different volumes or contain different number of copolymers. Furthermore, the shape of these WSCs is non-spherical in general. The formation of complex spherical packing phases requires the deformation of the polymeric domains from their natural state, {\it i.e.}, a sphere with a given size, into different non-spherical domains with different sizes. Any mechanisms favouring these domain deformations would be advantageous for the formation of the complex spherical packing phases \cite{li2017nonclassical,shi2021frustration,reddy2018stable}.

\begin{figure*}[htbp]
\centering
\includegraphics[width=7cm]{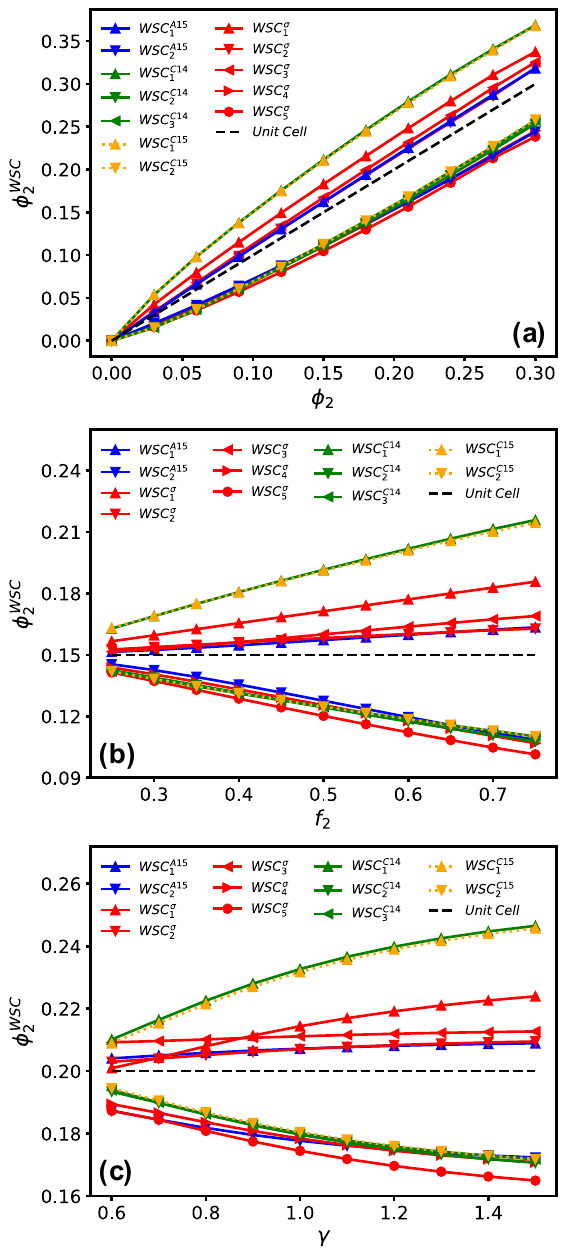}
\caption{$\text{A}_2\text{B}_2$ concentration ($\phi_{2}^{WSC}$), in different WSCs for the A15, C14, C15, and $\sigma$ phases, as a function of (a) $\phi_2$ (with fixed $f_{1}=0.2$, $f_{2}=0.7$, $\gamma_{1}=1.5$ and $\chi{N}=30$), (b) $f_2$ (with fixed $f_{1}=0.2$, $\gamma_{1}=1.5$, $\phi_{2}=0.15$, $\chi{N}=30$ and $\chi{N}=30$) and (c) $\gamma$ (with fixed $f_{1}=0.2$, $f_{2}=0.5$, $\phi_{2}=0.2$ and $\chi{N}=30$).}
\label{Inter1}
\end{figure*}

It has been proposed that the formation of the complex spherical packing phases in binary $\text{A}_1\text{B}_1$/$\text{A}_2\text{B}_2$ diblock copolymer blends is enhanced by the differential distributions of the two types of copolymers among the different WSCs. In particular, this inter-domain segregation of the different copolymers offers a mechanism to regulate the size of the spherical domains because the two diblock copolymers will occupy different volumes, $N/\rho_0$ and $\gamma N/\rho_0$, respectively. In order to demonstrate the non-uniform distribution, or local segregation, of the copolymers among the different WSCs quantitatively, we compute the concentration of the $\text{A}_2\text{B}_2$-copolymers ($\phi_{2}^{WSC}$) within the different WSCs. The calculated $\phi_{2}^{WSC}$ is plotted as a function of three molecular parameters, {\em i.e.} $\phi_2$, $f_2$ and $\gamma$, along certain phase paths in Figure~\ref{Inter1}(a), (b) and (c). 

One interesting and important observation from Figure~\ref{Inter1} is that the $\text{A}_2\text{B}_2$ concentrations $\phi_{2}^{WSC}$ in different WSCs deviate from the mean value , $\phi_2$, for all the complex spherical packing phases. This inter-domain segregation of diblock copolymers is especially apparent in Figure~\ref{Inter1}(a), where $\phi_{2}^{WSC}$ is plotted as a function of $\phi_{2}$. If there was no inter-domain segregation, all the curves should follow the straight line $\phi_{2}^{WSC}=\phi_{2}$. The deviations from this straight line indicate that the local concentration of the $\text{A}_2\text{B}_2$-copolymers in different WSCs becomes different from the average concentration. The WSCs could be roughly divided into two populations containing less or more $\text{A}_2\text{B}_2$-copolymers, corresponding to smaller and larger domains for the case with $\gamma=1.5$ shown in Figure~\ref{Inter1}(a). It is interesting to note that the smaller WSCs of the different ordered phases have similar values of $\phi_{2}^{WSC}$, whereas the $\phi_{2}^{WSC}$ of the larger WSCs are more scattered. A detailed examination of these curves reveals that the value of  $\phi_{2}^{WSC}$ is correlated with the coordination number (CN) of the WSCs. Specifically, the smallest values of $\phi_{2}^{WSC}$ are found for the WSCs with CN=12, whereas the largest values of $\phi_{2}^{WSC}$ are for the WSCs with CN=16. It is also interesting to observe that the $\phi_{2}^{WSC}$ values are approximately the same for the WSCs with the same CN, regardless of the phases. In particular, all the WSCs with the smallest CN=12 roughly have the same value of $\phi_{2}^{WSC}$. This grouping of $\phi_{2}^{WSC}$ according to the CN is also found when $\phi_{2}^{WSC}$ is plotted as a function of $f_2$ (Figure~\ref{Inter1}(b)) or $\gamma$ (Figure~\ref{Inter1}(c)). For the same ordered phase, the dispersity of the $\phi_{2}^{WSC}$ values among the different WSCs becomes larger when the molecular parameters ($\phi_2$, $f_2$, or $\gamma$) are increased. Namely, for a given value of $\phi_2$ a larger value of $f_2$ or $\gamma$ would enhance the inter-domain segregation. From these observations, it can be concluded that the formation of WSCs with different CN is correlated with the distribution of the $\text{A}_2\text{B}_2$-copolymers among the different WSCs. 

\begin{figure*}[htbp]
\centering
\includegraphics[width=7cm]{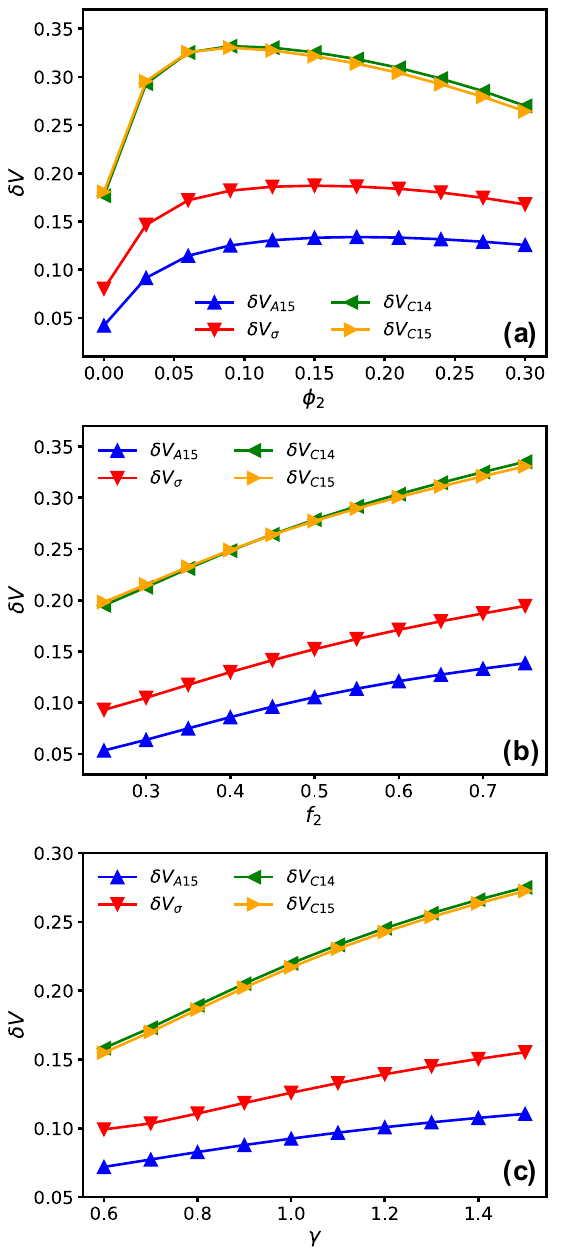}
\caption{The standard deviation of domain volume, $\delta{V}$, for the Frank-Kasper and Laves phases, as a function of (a) $\phi_2$, (b) $f_2$ and (c) $\gamma$. The other parameters are chosen as the same as those in Figure~\ref{Inter1}.}
\label{Inter2}
\end{figure*}

The difference between the maximum and minimum values of $\phi_{2}^{WSC}$, $\Delta\phi_{2}^{WSC}$, follows the order of $\Delta\phi_{2,A15}^{WSC}<\Delta\phi_{2,\sigma}^{WSC}<\Delta\phi_{2,C15}^{WSC}\approx\Delta\phi_{2,C14}^{WSC}$. This order of differences in $\phi_{2}^{WSC}$ is in agreement with the standard deviation of the WSC volumes of the different phases shown schematically in Figure~\ref{FK}, which have the ascending order of BCC/FCC/HCP(0.0) $<$ A15($\sim$0.0135316) $<$ $\sigma$($\sim$0.0421305) $<$ C14($\sim$0.0987861) $\lesssim$ C15($\sim$0.100634). It is noted that the volume deviation for the WSCs of the Laves phases is much larger than that of the A15 and $\sigma$ phases. In order to be quantitative, we have computed the standard deviation of the A-domain volumes ($\delta{V}$) for different phases and plotted the results as a function of $\phi_2$, $f_2$ and $\gamma$ in Figure~\ref{Inter2}. It is obvious that $\delta{V}>0$ for all the complex spherical phases and the $\delta{V}$ for the Laves phases is always the greatest in the whole parameter range shown in Figure~\ref{Inter2}. 

One interesting behaviour seen in Figure~\ref{Inter2}(a) is that the $\delta{V}$ is not a monotonically increasing function of $\phi_2$. Specifically, when $\phi_2$ is beyond certain critical value, $\delta{V}$ becomes saturated and then starts to decrease slightly. This suggests that an optimal ratio of the concentrations of longer and shorter chains exists that maximizes the benefit from raising the size difference through inter-domain segregation, exceeding which there is no further benefit upon adding the longer chains. The optimal ratio should be structure-dependent. Phases that have similar property of WSCs, such as C14 and C15 phases, have similar critical $\phi_2\sim0.1$ while $\sigma$ and A15 phases with smaller $\delta{V}$ have higher critical $\phi_2$. Another interesting observation is that $\delta{V}$ increases as increasing $f_2$ and $\gamma$. The mechanism of this tendency could be elucidated in the following discussion of intra-domain segregation.

It is worth to mention that the behaviour of $\phi_{2}^{WSC}$ and $\delta{V}$ at low $\phi_2$ (Figures~\ref{Inter1},~\ref{Inter2}) is similar to that of the AB/A binary blends where the A-homopolymers act as fillers localized in the central region of the A-domains \cite{xie2021giant}. In the case of $\text{A}_1\text{B}_1/\text{A}_2\text{B}_2$ block copolymer blends, the long A$_2$-blocks of the A$_2$B$_2$-copolymers play a similar role to occupy the central region of the A-domains. The differential segregation of these additives provides a mechanism to regulate the sizes of the domains, thus stabilizing the complex spherical packing phases, such as the Laves C14 and C15 phases, with large domain size dispersity. However, there is a noticeable difference between these two types of additives. In the case of AB/A binary blends, the A-homopolymers are localized at the central region of the domains to form a core composed mostly of A-homopolymers. On the other hand,  in $\text{A}_1\text{B}_1/\text{A}_2\text{B}_2$ diblock copolymer blends, there is a radial segregation of the long and short A-blocks forming a ``core-shell" structure, as will be illustrated in the next section.

\subsubsection{Intra-domain Segregation}

In contrast to the case of  binary AB/A diblock copolymer/homopoplymer blends, in which the A-homopolymers mainly act as fillers localizing inside the core of each domain, the added $\text{A}_2\text{B}_2$ diblock copolymers in the binary $\text{A}_1\text{B}_1/\text{A}_2\text{B}_2$ block copolymer blends will distribute across the A- and B-domains because of the architecture of the $\text{A}_2\text{B}_2$-copolymers. Specifically, the AB-junctions of the $\text{A}_2\text{B}_2$-copolymers will be localized at the AB-interfaces of the structure. In this case the $\text{A}_2\text{B}_2$-copolymers can act as fillers such that the long A-blocks will be stretched into the centre of the A-domain and, at the same time, as co-surfactants due to the localization of the AB-junctions. Therefore, the distribution of the $\text{A}_2\text{B}_2$-copolymers in the system possesses interesting internal patterns, in the form of radial segregation of the A-blocks and lateral segregation of the copolymers on the AB-interfaces, that will play a dual role as fillers to regulate the size of the polymeric domains and as co-surfactants to modify the interfacial properties \cite{liu2016stabilizing}. These two effects provide efficient mechanisms to stabilize the complex spherical packing phases. In what follows, we provide a detailed quantitative analysis of the internal structure of domains using the SCFT results and examine the effects of $\phi_2$, $f_2$ and $\gamma$ on the such structures. 

\begin{figure*}[htbp]
\centering
\includegraphics[width=10cm]{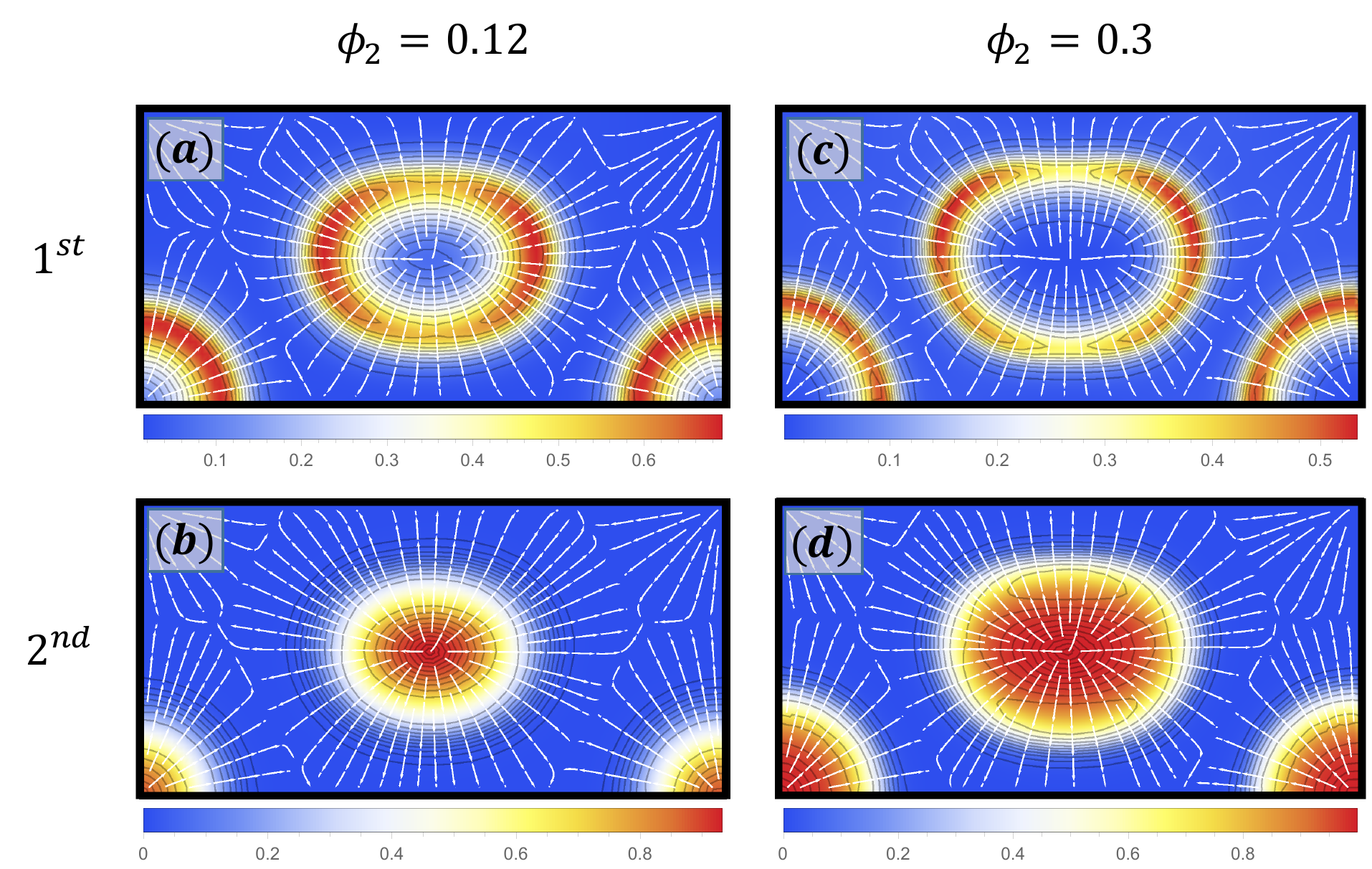}
\caption{The A-segment density and bond orientation distributions for $\text{A}_1\text{B}_1$ ((a) and (c)), and $\text{A}_2\text{B}_2$ ((b) and (d)), with $\phi_2=0.12$ ((a) and (b)) and $\phi_{2}=0.30$ ((c) and (d)). The other fixed parameters are $f_{1}=0.2$, $f_{2}=0.7$, $\gamma_{1}=1.5$ and $\chi{N}=30$. The plot is for the CN=14 domain of an equilibrated A15 structure.}
\label{Intra11}
\end{figure*}

\begin{figure*}[htbp]
\centering
\includegraphics[width=10cm]{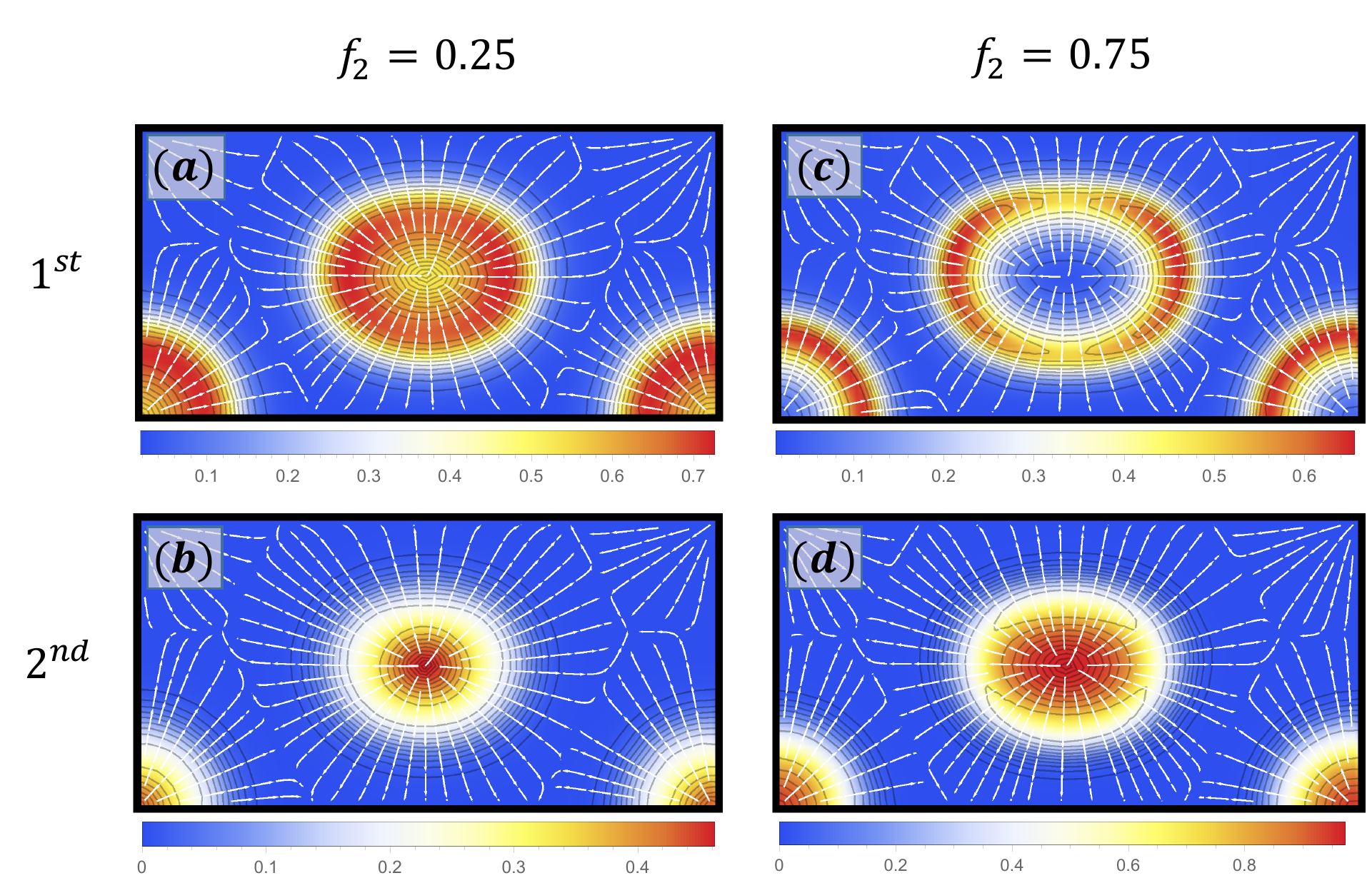}
\caption{The A-segment density and bond orientation distributions for $\text{A}_1\text{B}_1$ ((a) and (c)) and $\text{A}_2\text{B}_2$ ((b) and (d)), with $f_2=0.25$ ((a) and (b)) and $f_{2}=0.75$ ((c) and (d)). The other fixed parameters are $f_{1}=0.2$, $\gamma_{1}=1.5$, $\phi_{2}=0.15$ and $\chi{N}=30$. The plot is for the CN=14 domain of an equilibrated A15 structure.}
\label{Intra12}
\end{figure*}

\begin{figure*}[htbp]
\centering
\includegraphics[width=10cm]{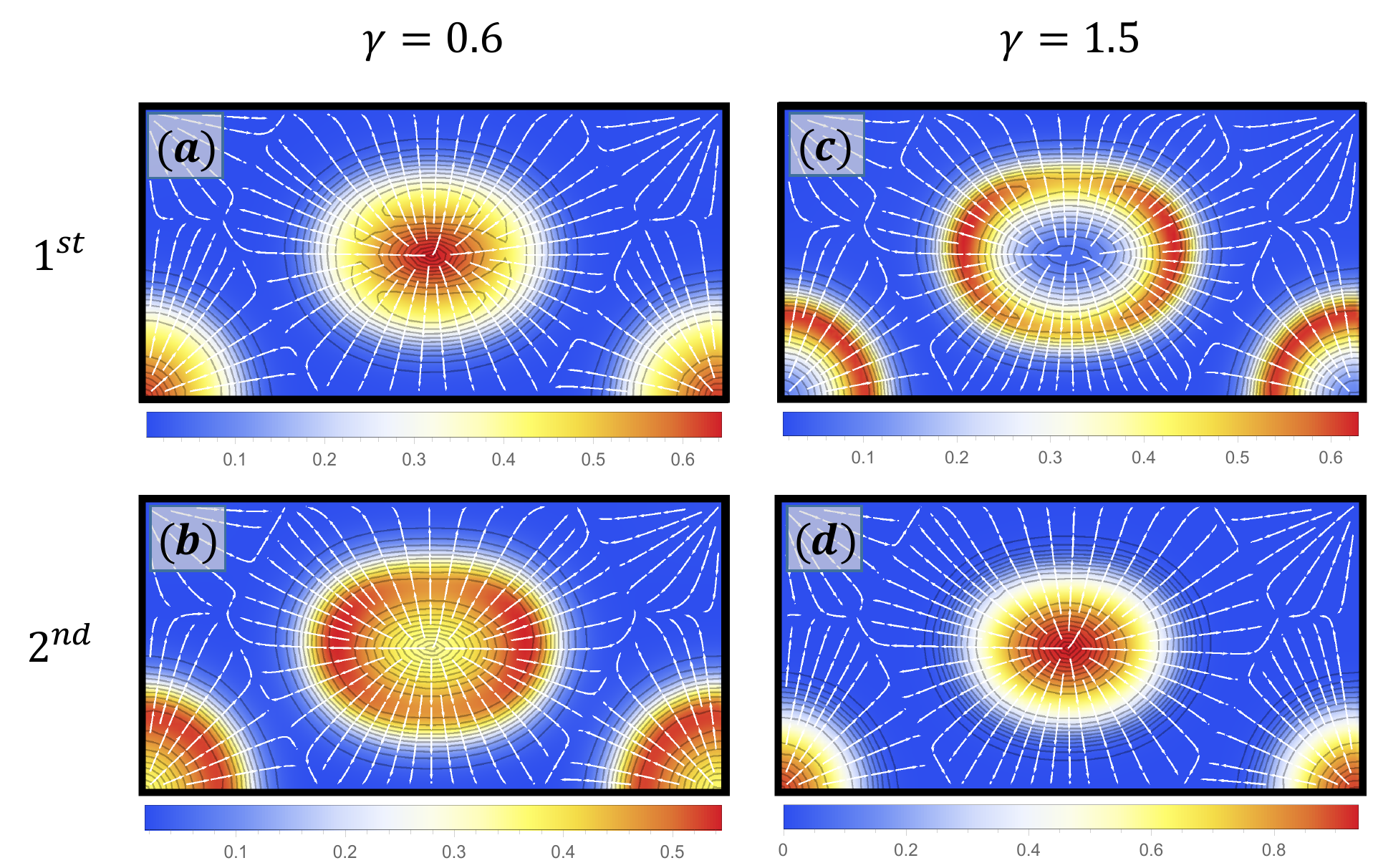}
\caption{The A-segment density and bond orientation distributions for $\text{A}_1\text{B}_1$ ((a) and (c)) and $\text{A}_2\text{B}_2$ ((b) and (d)), with $\gamma=0.6$ ((a) and (b)) and $\gamma=1.5$ ((c) and (d)). The other fixed parameters are $f_{1}=0.2$, $f_{2}=0.5$, $\phi_{2}=0.2$ and $\chi{N}=30$. The plot is for the CN=14 domain of an equilibrated A15 structure.}
\label{Intra13}
\end{figure*}

Due to the separation of the A- and B-blocks, the A-blocks are stretched from the AB-interface into the A-domains. Because the $\text{A}_1$- and $\text{A}_2$-blocks usually have different lengths, their degrees of stretching are different, resulting in a radial segregation of the two different A-blocks to form a ``core-shell'' structure. The segregation of the A-blocks or the formation of  the ``core-shell" structure could be revealed by the density distribution of the A-segments of the two copolymers. As an example, we plot the density and average bond orientation distributions of the A-segments for the CN=14 domain of an equilibrated A15 structure in Figures~\ref{Intra11}, \ref{Intra12} and \ref{Intra13}. The average bond orientation distribution was proposed by Prasad {\it et al.} to depict the average extension and orientation of the A-backbones \cite{prasad2017intradomain}. The inclusion of this quantity in the Figures provides information about the average stretching of the block copolymers. The existence of a radial separation of the $\text{A}_1$- and $\text{A}_2$-blocks is clearly visible in the density plots shown in Figures~\ref{Intra11}, \ref{Intra12} and \ref{Intra13}. Specifically, the longer $\text{A}$-blocks are stretched to reach the centre of the domain to form a core, whereas the shorter $\text{A}$-blocks are distributed near the AB-interface to form a shell. It is also interesting to observe that the orientation of the polymeric segments is perpendicular to the interface, especially in the region near the interface. 

For the $\text{A}_{i}\text{B}_{i}$ ($i=1,2$) diblock copolymers, the length of the A-blocks is specified by $N_{i,A}=f_{i}\gamma_{i}N$, thus the length ratio of the two A-blocks is $N_{2,A}/N_{1,A}=\gamma f_2/f_1$. For the case shown in Figure~\ref{Intra11}, we have $\gamma=1.5$, $f_1=0.2$ and $f_2=0.7$, thus $N_{2,A}/N_{1,A}=5.25$ so that the $\text{A}_2$-block is much longer than the $\text{A}_1$-block. In this case a well-defined core-shell structure is found in Figure~\ref{Intra11}, in which the the longer A$_2$-blocks form the core of the domain and the shorter A$_1$-blocks form a shell enclosing the core. This core-shell structure is found for different values of $\phi_2=0.12$ and $0.30$. The size of the domain increases significantly when the concentration of the the longer $\text{A}_{2}\text{B}_{2}$ diblock copolymers is increased from $0.12$ to $0.30$. At the same time, noticeable deformation of the domain occurs such that the overall domain shape approaches to that of the WSC. For a polymeric domain formed from monodisperse diblock copolymers, enlarging its size would result in extra stretching of the A-blocks. Such unfavourable stretching is avoided in the case of binary $\text{A}_1\text{B}_1/\text{A}_2\text{B}_2$ diblock copolymer blends because the longer A-blocks can be stretched towards the centre of the domain so that no excess stretching is required. Furthermore, the concentration of the longer diblock copolymers acts as a control parameter to regulate the domain size as illustrated by Figure~\ref{Intra11}. The inter-domain segregation shown in the Figure~\ref{Inter1} provides a mechanism to form polymeric domains with different sizes, which in turn pack into the complex spherical phases. 

The effects of $f_2$ and $\gamma$ on the ``core-shell" structure are examined in Figures~\ref{Intra12} and \ref{Intra13}, respectively. An obvious observation is that a well-defined ``core-shell" structure requires a large value of $N_{2,A}/N_{1,A}=\gamma f_2/f_1$. In the case shown in Figure~\ref{Intra12} with $\gamma=1.5$ and $f_1=0.2$, we have $N_{2,A}/N_{1,A}=1.875$ and $N_{2,A}/N_{1,A}=5.625$ for $f_2=0.25$ and $0.75$, respectively. As shown in Figure~\ref{Intra12} and in agreement with the expectation from the $N_{2,A}/N_{1,A}$ values, a well-defined core-shell structure is observed for $f_2=0.75$ whereas the shell becomes quite thick for $f_2=0.25$. Besides the length of the A-blocks, the overall length of the diblock copolymers could have significant effects on the radial segregation of the A-blocks inside a domain. This effect is illustrated by the case shown in Figure~\ref{Intra13} with $f_1=0.2$ and $f_2=0.5$, the ratio of $N_{2,A}/N_{1,A}$ is given by $N_{2,A}/N_{1,A}=1.5$ and $3.75$ for $\gamma=0.6$ and $1.5$, respectively. Although the $\text{A}_2$-block is slightly longer than the A$_1$-block for $\gamma=0.6$, Figure~\ref{Intra13}(a) and (b) reveal that the A$_2$-blocks form a rather thick shell whereas the A$_1$-blocks mostly localized at the core. This result suggests that in the binary blends of $\text{A}_1\text{B}_1$/$\text{A}_2\text{B}_2$ diblock copolymers, the copolymers with much shorter overall chain length tend to segregate at the interface even though their A-block is slightly longer. Based on these observations, it can be concluded that the formation of the complex spherical packing phases via the ``core-shell" structure requires that the additive $\text{A}_2\text{B}_2$-copolymers have a large  $N_{2,A}/N_{1,A}$ ratio {\em and} a chain length longer than, or at least comparable to, that of the $\text{A}_1\text{B}_1$-copolymers. As an example, none of the complex spherical phases are stable in the phase diagrams shown in Figure~\ref{PD2}(a) even though $N_{2,A}/N_{1,A}=\gamma{f_2}/f_1=1.25>1$, presumably due to the significantly shorter $\text{A}_{2}\text{B}_{2}$-copolymers.

\begin{figure*}[htbp]
\centering
\includegraphics[width=10cm]{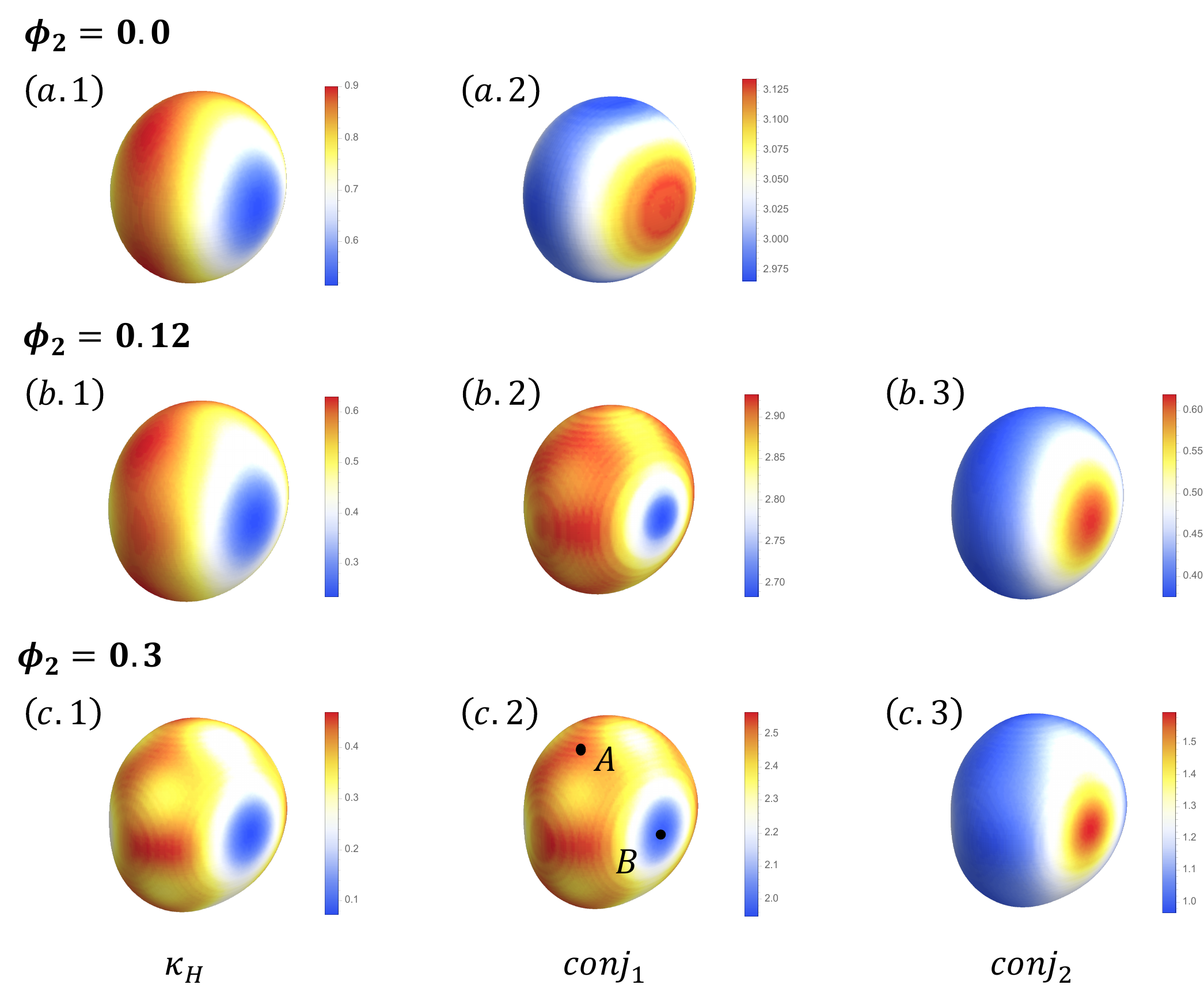}
\caption{Interfacial mean curvature ($\kappa_H$) and A$_i$B$_i$ junction distributions ($conj_i$, $i=1,2$) projected on the interface for $\phi_2=0$ (a), $0.12$ (b) and $0.30$ (c) and fixed $f_{1}=0.2$, $f_{2}=0.7$, $\gamma_{1}=1.5$ and $\chi{N}=30$. The plot is for the CN=14 domain of an equilibrated A15 structure.}
\label{Intra21}
\end{figure*}

\begin{figure*}[htbp]
\centering
\includegraphics[width=10cm]{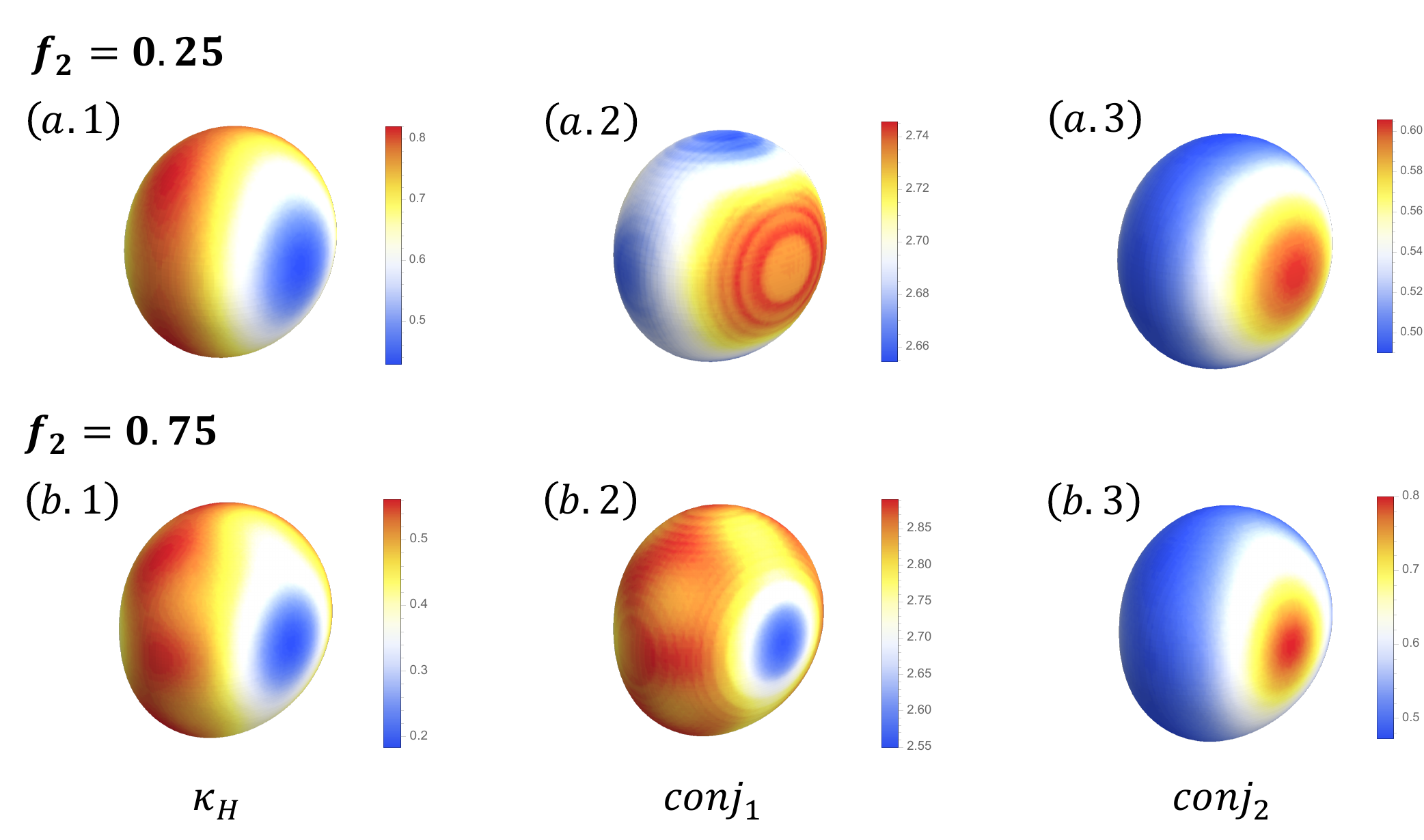}
\caption{Interfacial mean curvature ($\kappa_H$) and A$_i$B$_i$ junction distributions ($conj_i$, $i=1,2$) projected on the interface for $f_2=0.25$ (a) and $0.75$ (b), with fixed $f_{1}=0.2$, $\gamma_{1}=1.5$, $\phi_{2}=0.15$ and $\chi{N}=30$. The plot is for the CN=14 domain of an equilibrated A15 structure.}
\label{Intra22}
\end{figure*}

\begin{figure*}[htbp]
\centering
\includegraphics[width=10cm]{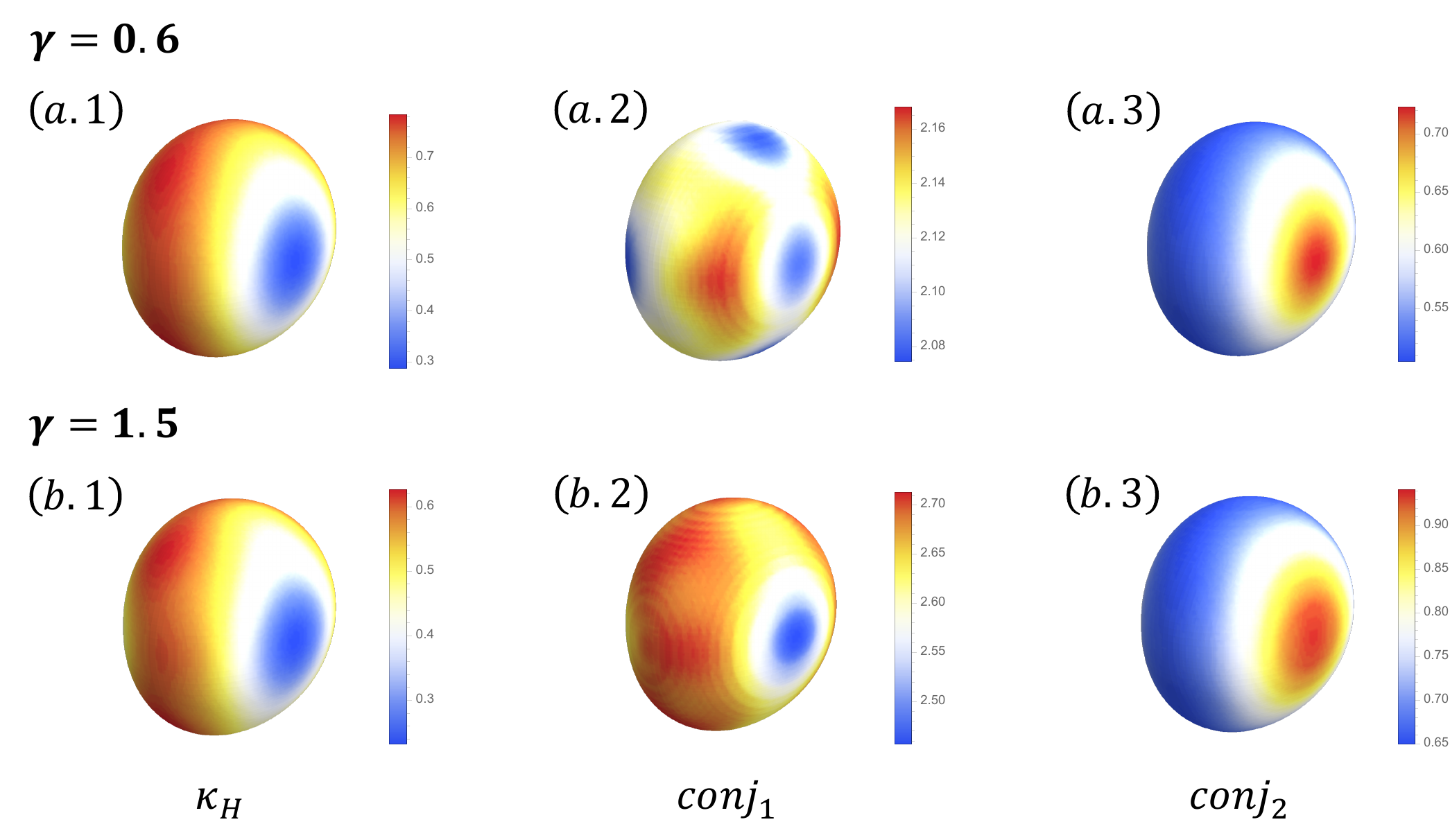}
\caption{Interfacial mean curvature ($\kappa_H$) and A$_i$B$_i$ junction distributions ($conj_i$, $i=1,2$) projected on the interface for $\gamma=0.6$ (a) and $1.5$ (b), with fixed $f_{1}=0.2$, $f_{2}=0.5$, $\phi_{2}=0.2$ and $\chi{N}=30$. The plot is for the CN=14 domain of an equilibrated A15 structure.}
\label{Intra23}
\end{figure*}

Due to the broken rotational symmetry in a crystalline structure, the WSCs assume the shape of a polyhedron. The AB-interface in the spherical packing phases would assume a non-spherical shape resembling the shape of the WSCs. On the other hand, the native shape of the polymeric domains is spherical with a uniform interfacial curvature. In the case of block copolymer blends, different diblock copolymers would have different preferred interfacial curvatures. It is therefore expected that the diblock copolymers would have an inhomogeneous distribution on the AB-interface and the distribution is coupled with the interfacial curvature. The interfacial distribution of the diblock copolymers could be revealed by the density of the AB-junction points of the two copolymers projected on the interface defined by the $\phi_{A}(\vec{r})=\phi_{B}(\vec{r})$ isosurface. To facilitate the comparison between the distribution of AB-junctions and the interfacial curvature, we also compute the mean curvature of the interface, $\kappa_{H}$, based on level set method \cite{albin2016computational}. As an example, the mean-curvature and interfacial AB-junction distributions have been computed for the CN=14 domain of an equilibrated A15 structure and are shown in Figures~\ref{Intra21}, \ref{Intra22} and \ref{Intra23} for three sets of molecular parameters.

The plots shown in Figures~\ref{Intra21}, \ref{Intra22} and \ref{Intra23} clearly reveal the existence of a curvature driven lateral segregation of the AB-junctions on the AB-interfaces. It is interesting to observe that a curvature driven lateral segregation of the AB-junctions occurs, {\it albeit} with a very small amplitude, even for the neat A$_1$B$_1$-copolymers as shown in Figure~\ref{Intra21}(a) where a relatively higher density of AB-junctions is found in the area with a smaller curvature. When the longer and less asymmetric A$_2$B$_2$-copolymers are added to the system, a curvature driven segregation of the A$_1$B$_1$- and A$_2$B$_2$-copolymers on the AB-interface takes place, such that the A$_2$B$_2$-copolymers are localized in the area with low curvature whereas the A$_1$B$_1$-copolymers concentrate in the area with high curvature. Specifically, as shown in Figure \ref{Intra21}, the pattern of the mean curvature and junction distributions coincide with the shape of the WSC, especially when $\phi_2$ is large. For example, the points $A$ and $B$ in Figure~\ref{Intra21}(c.2) represent roughly a local maximum and minimum of the $\text{A}_{1}\text{B}_{1}$-junction density and they correspond to the vertex and face of the WSC, respectively. Because the vertices and faces of the WCS correspond to high and low interfacial curvature areas, the accumulation and depletion of the $\text{A}_{1}\text{B}_{1}$-junctions in these areas are expected. 

As shown in Figure~\ref{Intra22}, the A-volume fraction $f_2$ of the A$_2$B$_2$-copolymers affects significantly the interfacial segregation of the diblock copolymers. When $f_2=0.25$ which is slightly larger than $f_1=0.2$, the interfacial segregation behaviour is very similar to that for the neat A$_1$B$_1$-copolymers (Figure~\ref{Intra21}(a)) such that there is a weak accumulation of the A$_1$B$_1$-junctions at the low curvature area. On the other hand, when $f_2$ changes to $f_2=0.75$, the interfacial segregation pattern is inverted such that the A$_1$B$_1$-junctions are depleted in the low curvature area. Similar to the formation of the ``core-shell" structure, the interfacial segregation also requires simultaneously large $f_2$ and $\gamma$ as revealed by Figure \ref{Intra22} and \ref{Intra23}. Particularly, for the chains with a large $f_2$ but significantly smaller total chain length or smaller $\gamma$, the distribution of the A$_1$B$_1$-junctions are nearly constant on the interface as indicated by Figure \ref{Intra23} (a.2), where the variation of the junction density is extremely small as depicted on the scale bar.

\section{Conclusion}

In summary, we have systematically studied the formation and relative stability of complex spherical packing phases in binary blends composed of A$_1$B$_1$ and A$_2$B$_2$ diblock copolymers with different chain lengths and compositions by using the polymeric self-consistent field theory. A set of phase diagrams of the binary blends have been constructed, representing the phase behaviour of the system in a large parameter space. The effects of three molecular parameters, {\em i.e.} the concentration $\phi_2$, composition $f_2$ and relative chain length $\gamma$ of the A$_2$B$_2$ diblock copolymers, have been examined. Our results predict that the complex spherical packing phases, {\em i.e.} the Frank-Kasper A15 and $\sigma$ and the Laves C14 and C15 phases, can become stable equilibrium phases with proper choices of $f_2$ and $\gamma$. In particular, the theoretical results reveal that large values of $f_2$ and $\gamma$ are required simultaneously to stabilize the complex spherical packing phases. The phase diagrams could be used to predict the phase transition sequence for a given set of parameters. For example, for BCC-forming A$_1$B$_1$ diblock copolymers with $f_1=0.2$, the addition of A$_2$B$_2$, or increasing $\phi_2$ from 0 to 1, with $f_2=0.7$ and $\gamma=1.5$ is predicted to induce order-order phase transitions following the sequence of BCC $\rightarrow$ C14 $\rightarrow$ C15 $\rightarrow$ $\sigma$ $\rightarrow$ A15 $\rightarrow$ HEX $\rightarrow$ L $\rightarrow$ DG $\rightarrow$ HEX. The predicted phase behaviour is in agreement with available theoretical studies and experiments. More importantly, the theoretical results predict that the binary A$_1$B$_1$/A$_2$B$_2$ diblock copolymer blends provide an efficient and versatile platform to obtain complex spherical packing phases.

The mechanisms stabilizing the complex spherical packing phases have been explored by a detailed and quantitative analysis of the SCFT solutions. Specifically, the spatial distributions of the different polymeric species are used to establish correlations between structural formation and polymer segregation. A detailed examination of the $\text{A}_{2}\text{B}_{2}$-concentration in different Wigner-Seitz cells reveals that inter-domain segregation of the different diblock copolymers occurs in the blends, resulting in domains of different sizes. At the same time, an examination of the A-segment density and AB-junction distributions demonstrates that two types of intra-domain segregation take place. The radial segregation of the long and short A-blocks inside the A-domains results in a core-shell structure. The simultaneous stretching of the long and short chains enables the formation of large spherical domains. The lateral segregation of the AB diblock copolymers on the AB-interfaces releases the frustration of forming non-spherical cells. These mechanisms operate synergistically resulting in a large stable region for the complex spherical packing phases. 

The current study focused on the simplest binary blends of diblock copolymers, {\em i.e.}, blends containing $\text{A}_{1}\text{B}_{1}$ and $\text{A}_{2}\text{B}_{2}$ diblock copolymers. 
It is natural to expect that more complex binary blends composed of AB/CD diblock copolymers will exhibit more complex phase behaviours, thus providing more opportunities to form novel ordered phases, 
 {\em e.g.} other Frank-Kasper phases and quasicrystals.
Extension of the current study to more complex blends containing block copolymers is straightforward, however, care must be taken when one chooses the molecular parameters and possible ordered candidate phases. 

The results obtained in the current study provide a useful foundation for further investigation of more complex polymeric blends containing block copolymers. The mechanisms identified in the study will be helpful for the understanding of the formation of complex structures via polymer self-assembly. In particular, the mechanism of local segregation of distinct components in polymer blends provides a simple and effective route to access complex ordered phases, particularly the complex spherical packing phases.

\section{Acknowledgements}

We acknowledge delightful discussions on the phase behaviour of block copolymer blends with Frank Bates, Kevin Dorfman, Weihua Li, Meijiao Liu and Rob Wickham. This research was supported by the Natural Sciences and Engineering Research Council (NSERC) of Canada and was enabled in part by support provided by the facilities of SHARCNET (https://www.sharcnet.ca) and Compute Canada (http://www.computecanada.ca).


\end{document}